\documentclass{aa}

\usepackage{txfonts}
\usepackage{graphicx}
\usepackage{hyperref}

\def\msun{\hbox{M$_\odot$}}
\newcommand{\sfr}{\mathrm{SFR}}

\begin{document}

\title{The apparent Large Magellanic Cloud star cluster age gap}

\author{J.H.~Klos\inst{1}\thanks{\email{jonathan.klos@stud.uni-frankfurt.de}} \and A.E.~Piatti\inst{2,3}\thanks{\email{andres.piatti@fcen.uncu.edu.ar}}}

\institute{
Goethe-Universität Frankfurt, Institut für Mathematik, Robert-Mayer-Straße 10, 60325 Frankfurt am Main, Germany
\and
Instituto Interdisciplinario de Ciencias B\'asicas (ICB), CONICET-UNCuyo, Padre J. Contreras 1300, M5502JMA, Mendoza, Argentina
\and Consejo Nacional de Investigaciones Cient\'{\i}ficas y T\'ecnicas (CONICET), Godoy Cruz 2290, C1425FQB,  Buenos Aires, Argentina
}

\date{Received 26 January 2026 / Accepted 20 April 2026}

\abstract
{In the Large Magellanic Cloud (LMC), there have been very few clusters observed with ages between 4 and 11  Gyr. This phenomenon is sometimes referred to as the `LMC age gap'.}
{We constructed a model of the cluster age distribution aimed at reproducing this scenario.}
{We linked the star formation history to the cluster initial mass function via a power-law relation between maximum initial cluster mass and global star formation rate. Using a constant cluster-forming efficiency of 5\%, we obtained the cluster formation history. Applying a model of cluster mass loss calibrated using $N$-body simulations and an observational completeness limit, we computed the observable fraction of initially formed clusters. We were then able to model the cluster age distribution.}
{For a maximum initial cluster mass below $10^5\msun$ at a star formation rate of $1 \msun\mathrm{pc^{-2} Gyr^{-1}}$, our model reproduced the observed lack of clusters with ages between 4 and 11 Gyr. However, our model required a maximum initial mass at $1 \msun\mathrm{pc^{-2} Gyr^{-1}}$ of at least $2\cdot 10^5\msun$ to reproduce the population of ancient globular clusters. A linear change between maximum initial cluster mass relations from 8 to 12 Gyr reproduced the age gap to a satisfactory extent.}
{In our model, the age gap is a consequence of the star-forming history and current observational limits. The age gap corresponds to a period characterised by a lower star formation rate, whereby no clusters with an initial mass above approximately 2 to $5\cdot 10^5\msun$ were formed. In the present day, these clusters have become so faint that very few of them have been detected. The pattern of both young-and-bright and old-and-massive clusters being more easily detectable than clusters of intermediate ages might reflect a more general phenomenon and not necessarily one specific to the LMC.}


\titlerunning{LMC cluster age gap}  

\authorrunning{J.H.~Klos and A.E.~Piatti}           

\maketitle

\markboth{J.H.~Klos and A.E.~Piatti:}{LMC cluster age gap}

\section{Introduction}           

In the lifetime of the Large Magellanic Cloud (LMC), the period
between about 4 and 11 Gyr is known as the `LMC age gap',
characterised by a relatively small number of discovered star clusters. 
This is noteworthy as the star-forming history inferred from field stars is not commensurately small. The LMC age gap can therefore be characterised as an observational mismatch between cluster and field star population that needs to be resolved either by discovery of the `missing' clusters or by some mechanism explaining the low number of clusters observed.
 
During the last decades, different observational campaigns 
have searched unsuccessfully for new LMC age gap clusters 
\citep{dc1991,getal97,piatti2021d,piatti2025c}. Indeed, up to date, 
only four star clusters have reliable age estimates that undoubtedly place 
them in the LMC age gap, namely: ESO121-SC03 \citep[][]{mateoetal1986}, 
KMHK~1592 \citep[][]{piatti2022c}, KMHK~1762 \citep[][]{gattoetal2022}, 
and SL~2 \citep[][]{ferreiraetal2025}, respectively. Some additional
star cluster candidates have been reported, but their shallow and
small number photometry serves as a caveat against considering them in this 
age cluster group \citep{pieresetal2016,gattoetal2020,gattoetal2024}.

Besides the large observational efforts summarised above, very few
theoretical simulations have attempted to explain the existence 
of the LMC age gap. \citet{bekkietal2004} proposed that the age gap
is related to the most recent star cluster formation
(age $\la$ 4 Gyr) being triggered by strong tidal interactions between the LMC, 
the Small Magellanic Cloud (SMC) and the Milky Way. As for the LMC formation
and evolution,
\citet{pt1998} modelled the chemical evolution of the Magellanic Clouds using 
a bursting star formation model with an important formation event at around 2.0 Gyr. 
They considered the cluster age distribution as a tracer of star formation bursts 
alongside a continuous field star formation.
More recently, \citet{2013MNRAS.430..676B} modelled the LMC cluster-forming history using the star formation history of \citet{2009AJ....138.1243H} and theoretical estimates of cluster dissolution.
They found that to reproduce the LMC cluster age function (CAF), cluster lifetimes had to be 10 times lower than theoretically expected, and their model did not reproduce an age gap.
The relatively small number of known age gap clusters points to the need of improvements of 
LMC cluster formation models. 

To find a plausible explanation of the small number
of LMC clusters with ages in the age gap range, we started by focussing on the cluster
mass range of the LMC star cluster population older than 4 Gyr. 
To this respect, we included the 15 known ancient LMC globular clusters and the four
aforementioned LMC age gap clusters. 
We found that the former group spans cluster masses between $7.4 \cdot 10^4$ and $7.6 \cdot 10^5$ $\msun$ \citep{pm2018,piattietal2019}, 
while the latter group  spans masses from 180 $\msun$ up to $5\cdot 10^3$ $\msun$ \citep{mateoetal1986,metal14,piatti2022c,ferreiraetal2025,piattietal2025}.
Despite being subjected to mass loss due to stellar evolution, two-body
relaxation and tidal heating for much longer, the oldest star clusters are much more massive
than the younger ones. 
This finding raises the possibility that there could be a 
connection between periods with enhanced star formation rates (SFRs) and the formation 
of more massive star clusters. 
According to the integrated cloud-wide initial mass function theory, the mass of the most massive star cluster 
is positively proportional to the SFR, the SFR surface density of the cloud where it formed, 
as well as to the mass of the cloud and the column density \citep{zhouetal2025}.
Likewise, \citet{lietal2018} carried out simulations and found that the fraction of 
clustered star formation and maximum cluster mass increase with the SFR surface 
density. \citet{bereketal2023} performed logistic regressions using the SFR and 
the total stellar mass in the galaxy as predictors, and found that the SFR is the better 
predictor for the probability of hosting clusters and the total mass in the cluster system. 
When they compared their results to similar models for old globular clusters, they
concluded that star cluster formation was more abundant and more efficient at 
higher redshifts, likely because of the high gas content of galaxies at that time.

In this work, we present an LMC cluster formation model based on the 
above ideas that satisfactorily reproduces the LMC age gap. This model has enabled us to provide, 
for the first time, an explanation of this phenomenon. 
We first describe the relevant basic relations between SFR, cluster formation rate (CFR) and cluster initial mass function (CIMF) needed to model the CAF in 
Sect.~\ref{sec:relations}. 
In Sect.~\ref{sec:massloss}, we couple these relations to an analytic model of cluster 
mass loss due to \cite{Lamers2010} to obtain a model mapping the SFR history to the CAF.
We describe how we used the LMC SFR data derived by \cite{2020A&A...639L...3R} 
to compute the corresponding expected CAF. 
We discuss our results as well as the limitations of our model in Sect.~\ref{sec:results}. 
Our final conclusions are presented in Sect.~\ref{sec:conclusion}.

\section{Modelling relations for the CAF}
\label{sec:relations}

\subsection{CFR}

We considered the CFR $\Psi$ in terms of the number of clusters formed per unit time and the SFR as stellar mass formed per unit time.
If $\Gamma$ is the fraction of clustered star formation (i.e.~the fraction of the total SFR occurring within newly formed clusters), we then have
\begin{equation}
\label{eq:gamma}
\Gamma = \frac{\langle M \rangle \Psi}{\sfr}.
\end{equation}
Here, $\langle M \rangle$ is the mean initial cluster mass, given by 
\begin{equation}
\langle M \rangle = \int\limits_{M_\mathrm{min}}^{M_\mathrm{max}} M f(M) \mathrm{d}M,
\end{equation}
where $f(M)$ is the CIMF with upper and lower mass limits, $M_\mathrm{max}$ and $M_\mathrm{min}$, respectively.
From observations, the dependence of $\Gamma$ on the SFR has been shown to be moderate at 
best \citep[see e.g.~Sect.~2.4 of][]{Krumholz2019}.
A constant value of $\Gamma$ is consistent with observations, and the effect of a variable $\Gamma$ on the CAF is comparatively weak.
Thus, we took $\Gamma$ to be constant, and used $\Gamma=0.05$ as a fiduciary value in the model, noting that the shape of the resulting CAF does not depend on the value of $\Gamma$.
The CFR can then be obtained from Eq.~(\ref{eq:gamma}) by rearranging it to $\Psi = \Gamma\cdot \sfr / \langle M \rangle$.

\subsection{CIMF}
\label{sec:cimf}

The CIMF we used takes the power-law form of
\begin{equation}
f(M) = k M^{-a},
\end{equation}
for $\alpha > 1$.
In particular, we considered the case of $\alpha=2$, where $k = \frac{M_\mathrm{max} M_\mathrm{min}}{M_\mathrm{max} - M_\mathrm{min}}$ and 
\begin{equation}
\langle M \rangle = k \ln \frac{M_\mathrm{max}}{M_\mathrm{min}}.
\end{equation}

As discussed above, $M_\mathrm{max}$ depends on the SFR density in the cluster-forming environment.
Following, for instance, ~\cite{2012MNRAS.420..340K}, we introduced a power-law scaling of $M_\mathrm{max}$ with the SFR taking the form of
\begin{equation}
\label{eq:max_mass}
M_\mathrm{max} = M_0 \left(\frac{\sfr}{\sfr_0} \right)^\beta,
\end{equation}
which we refer to as the maximum initial cluster mass (MICM) relation. The value of $\beta$ for a given environment and measurement of the SFR will depend, in addition to the star-forming processes themselves, on the spatial resolution of the SFR measurement and how strongly concentrated star formation is.
Regarding the relation for $M_\mathrm{max}$, \cite{2012MNRAS.420..340K} considered the work of \cite{2008MNRAS.390..759B}, who simulated the relation between CFR and maximum magnitude of young massive clusters for different CIMF cut-offs and compared this to a relation between maximum GC magnitude and global SFR observed in high-SFR galaxies.
Reviewing \cite{2008MNRAS.390..759B}, 
we found that their results may be compatible with values of $\beta$ up to 1.

In principle, $M_\mathrm{min}$ can also be expected to vary with the SFR.
However, the effect of this on the CAF is likely small, as the least massive clusters are expected to quickly dissolve.
As such, we modelled $M_\mathrm{min}$ as constant with value $M_\mathrm{min} = 100\msun$.

\subsection{CAF}
\label{sec:caf}

The CAF $\eta(t)$ as the present number of clusters at age $t$ is given by
\begin{equation}
\eta (t)  = f_\mathrm{surv}(t) \Psi(t) = f_\mathrm{surv}(t) \frac{\Gamma \cdot \sfr}{\langle M \rangle},
\end{equation}
where $ f_\mathrm{surv}(t)$ is the surviving 
fraction of clusters as a function of time, dependent on cluster mass loss. 
It implicitly depends on the SFR through the CIMF.

In practice, the low-mass end of the cluster population is not completely observed.
Therefore, we introduced $\eta_\mathrm{obs} (t)= f_\mathrm{obs}(t) \Psi(t)$ for the observable CAF, where $f_\mathrm{obs}(t)$ is the observable fraction of clusters, determined by both cluster mass loss and observational sensitivity.

For real cluster data, the CAF can be constructed either directly from all observed clusters or after applying a cut in mass or luminosity to account for observational incompleteness.
In the latter case, applying the same correction for incompleteness to the model CAF is necessary to compare modelled and observed cluster counts.
In the former case, the completeness function in age and mass must be modelled to obtain $f_\mathrm{obs}(t)$.
Incompleteness at low cluster masses is another reason why the choice of $M_\mathrm{min}$ has little impact on the observable CAF in our model.
As we discuss in Sect.~\ref{sec:fractions}, we used a mass cut to compare observed and modelled cluster counts. A comparison of $\Psi(t),\,\eta(t),$ and $\eta_\mathrm{obs} (t)$ in our model can be found in Fig.~\ref{fig:model}.

\section{Modelling cluster mass loss}
\label{sec:massloss}

To compute $ f_\mathrm{surv}(t)$ and $f_\mathrm{obs}(t)$, a model of cluster mass loss is needed.
For this, we used the description of \cite{Lamers2010}, 
which includes mass loss due to stellar evolution both directly (stellar mass loss) and indirectly (induced mass loss due to stars becoming unbound), as well as dynamical relaxation in the tidal field of the host galaxy.
The parameters of mass loss are computed by empirical relations obtained via N-body simulations for clusters with initial King parameters of $W_0=5$ and $W_0=7$. 
We refer to \cite{Lamers2010} for the full details (see in particular Appendix A therein) and we give only a brief overview here. We used the parameters for initially Roche-lobe filling clusters with $W_0=7$, but we note that for our model, the resulting CAF for $W_0=5$ is in general agreement within 2\%.

While this description does not include contributions from giant molecular cloud (GMC) encounters, we chiefly considered long-lived, massive LMC clusters. We argue that for these, GMC encounters do not affect mass loss significantly.

In brief, cluster mass loss is given via
\begin{equation}
\label{eq:dmdt}
\frac{\mathrm{d}m}{\mathrm{d}t} = \left( \frac{\mathrm{d}m}{\mathrm{d}t} \right)_\mathrm{ev} + \left( \frac{\mathrm{d}m}{\mathrm{d}t} \right)_\mathrm{dis} 
\end{equation}
where $\left( \frac{\mathrm{d}m}{\mathrm{d}t} \right)_\mathrm{ev}$ is the mass loss from stellar evolution and $ \left( \frac{\mathrm{d}m}{\mathrm{d}t} \right)_\mathrm{dis} $ the mass loss from tidal dissolution.

\subsection{Stellar evolution}

Mass loss from stellar evolution is modelled as 
\begin{equation}
\left( \frac{\mathrm{d}m}{\mathrm{d}t} \right)_\mathrm{ev} = - m_\mathrm{lum}(t) (1 + f_\mathrm{ind} (t)) \frac{\mathrm{d}\mu_\mathrm{ev}(t)}{\mathrm{d}t},
\end{equation}
where $m_\mathrm{lum}(t)$ is the total mass of stars that are still evolving, 
$ f_\mathrm{ind} (t)= f_\mathrm{ind}^\mathrm{max} f_\mathrm{delay}(t)$ is the fraction of induced mass loss, 
and $\mu_\mathrm{ev}$ is the fraction of initial mass remaining after stellar evolution.
\cite{Lamers2010} approximated $\mu_\mathrm{ev}$ as well as the production of stellar remnants using fourth-order polynomials in $\log (t /  \mathrm{Myr} )$ 
and give tables for metallicities $Z = 0.0004,\, 0.001,\, 0.004,\, 0.008$, and $0.02$. 
Depending on what fraction of black holes, neutron stars, and white dwarfs are removed by initial kicks, $m_\mathrm{lum}(t)$ and $\mu_\mathrm{ev}(t)$ must be adjusted.
We used $Z= 0.004$ as fiduciary value for the LMC and retained all stellar remnants.

The fractions $f_\mathrm{ind}^\mathrm{max}$ and $f_\mathrm{delay}(t)$ are described by empirical relations depending on the cluster initial mass, $M$, 
the crossing time at the tidal radius, $t_\mathrm{cr}(r_\mathrm{t})$, and a characteristic mass loss timescale, $t_0$.

The mass loss timescale, $t_0$, is given by 
\begin{equation}
\label{eq:t0}
t_0 = t^N_\mathrm{ref} \left(\frac{1}{\bar{m}^\gamma} \right)\left(\frac{R}{8.5 \mathrm{kpc}} \right)\left(\frac{v}{220 \mathrm{km \, s^{-1}}} \right).
\end{equation}
Here, $R$ is the orbital radius, $v$ the orbital velocity, $\bar{m}$ the mean stellar mass, and $ t^N_\mathrm{ref}$ and $\gamma$ 
are empirical parameters that depend on the cluster density profile and whether the cluster is in the pre- or post-core-collapse phase. We note that we considered only circular orbits and thus have dropped the terms depending on orbit eccentricity.

\subsection{Tidal dissolution}

Mass loss from tidal dissolution is given as
\begin{equation}
\label{eq:tidal}
\left( \frac{\mathrm{d}m}{\mathrm{d}t} \right)_\mathrm{dis} = - f_\mathrm{delay}(t) \frac{m^{1 - \gamma}}{t_0},
\end{equation}
using the quantities introduced above. 
While mass loss from stellar evolution scales linearly with cluster mass, mass loss from tidal dissolution has sub-linear scaling that follows from $\gamma > 0$ in Eq.~(\ref{eq:tidal}). 
Tidal forces have a stronger relative effect on less massive clusters, causing them to lose mass quickly, while more massive clusters are more resistant to tidal dissolution and lose a smaller fraction of their mass in the same time frame. 
As a result, relative differences in initial masses grow as clusters evolve.

\subsection{LMC tidal parameters}

As can be seen from Eq.~(\ref{eq:t0}), \cite{Lamers2010} published model parameters scaled to solar tidal conditions, but they also provided relations to rescale to arbitrary tidal environments. 
To compute the cluster tidal parameters for LMC clusters, we considered $I(R) = M_g(R)/R^3$, the cluster's tidal index for $M_g(R)$ representing the mass of the host galaxy enclosed within a radius, $R$.
From \cite{2025MNRAS.537.1586P},
we found that LMC clusters populate the disc at orbital radii of 100~pc to 20~kpc and estimated a piece-wise power law for the tidal index in the LMC by
\begin{equation}
\log (I(R) / \msun \mathrm{kpc}^{-3} )\approx 9.3 + k_\pm \times \log ({R}/{R_0})
,\end{equation}
with $R_0 = 2.5\mathrm{kpc}$ and $k_- = -1.2 $ for $R < R_0$, $ k_+ = -2.3 $ for $R>R_0$.

This allowed us to compute tidal parameters $v$, $r_\mathrm{t}$, and $t_\mathrm{cr}(r_\mathrm{t})$ for a cluster at given orbital radius, expressed as
\begin{equation}
v = \sqrt{G\, I(R) R^3} ,\quad r_\mathrm{t} = \left(\frac{m}{2I(R)}\right)^\frac{1}{3}  ,\quad t_\mathrm{cr}(r_\mathrm{t}) = \frac{2}{\sqrt{G\, I(R)}},
\end{equation}
where $G$ is the gravitational constant.
The cluster mass as a function of time, $m(t)$, was then obtained by numerical integration of the mass loss rate from Eq.~(\ref{eq:dmdt}). 
We did this using a standard fourth-order Runge-Kutta scheme with a time step of 1 Myr.

\subsection{Surviving and observable fraction}
\label{sec:fractions}

We used $m(t)$ to compute the 
surviving fraction $f_\mathrm{surv}(t)$.
As cluster mass loss depends on the orbital radius, we have
\begin{equation}
f_\mathrm{surv}(t;R) = \int_{M_\mathrm{dis}(R,t)}^{M_\mathrm{max}(t)} f(M)\,\mathrm{d}M
\end{equation}
for the surviving fraction at orbital radius, $R$. 
We note that $M_\mathrm{max}$ varies with $t$ depending on $\sfr (t)$.
The integral's lower limit of $M_\mathrm{dis}(R,t)$ gives the minimum initial mass for a cluster at a radius, $R$, to survive past age, $t$. 
In other words, for a cluster at $R$ with initial mass $M = M_\mathrm{dis}(R,t)$, the mass at age $t$ is $m(t) = m_\mathrm{dis}$, the lower mass limit below which a cluster is considered to be dissolved.
We used $m_\mathrm{dis} = 100\msun$.

To correct for observational incompleteness when comparing to observed cluster counts, we considered only clusters above a certain observational limit in age and mass.
The assumption is that all clusters above this limit are observed, while clusters below the limit are observed only incompletely.
Following \cite{2014MNRAS.437.1646P}, 
we set 
\begin{equation}
\log ( m_\mathrm{lim} / \msun) = 1.8 \times \log (t/\mathrm{yr}) -12.8
\end{equation}
as an observational completeness limit for LMC clusters older than 1~Gyr.
The observable fraction, $f_\mathrm{obs}(t)$, then depends on the observational completeness limit, $m_\mathrm{lim}(t)$, in terms of mass as a function of the cluster age.
At a time, $t$, for given $R$, we computed the minimum initial cluster mass, $M_\mathrm{lim}(R,t)$, such that the corresponding present-day mass, $m(t)$, is greater than $m_\mathrm{lim} (t)$. 
$M_\mathrm{lim}(R,t)$ is then the observational completeness limit in initial mass at a radius, $R$, and age, $t$, 
such that 
\begin{equation}
f_\mathrm{obs}(t;R) = \int_{M_\mathrm{lim}(R,t)}^{M_\mathrm{max}(t)} f(M)\,\mathrm{d}M,
\end{equation}
which gives the observable fraction at orbital radius $R$. 

Assuming that clusters are homogeneously distributed in the disc between radii $R_\mathrm{min}$ and $R_\mathrm{max}$, we then averaged over $R$ as
\begin{equation}
\begin{split}
f_\mathrm{obs}(t)&= \frac{1}{R_\mathrm{max}^2 - R_\mathrm{min}^2} \int\limits_{R_\mathrm{min}}^{R_\mathrm{max}} f_\mathrm{obs}(t;R) 2R \,\mathrm{d}R \\&= \frac{1}{R_\mathrm{max}^2 - R_\mathrm{min}^2} \int\limits_{R_\mathrm{min}}^{R_\mathrm{max}} \left( \int_{M_\mathrm{lim}(R,t)}^{M_\mathrm{max}(t)} f(M)\,\mathrm{d}M \right) 2R \,\mathrm{d}R.
\end{split}
\end{equation}
This gives the global observable fraction $f_\mathrm{obs}(t)$ for the LMC.
The global surviving fraction $f_\mathrm{surv}(t)$ can be computed from $f_\mathrm{surv}(t;R)$ by an analogous integral.

\begin{figure}
\center
\includegraphics[scale=0.6]{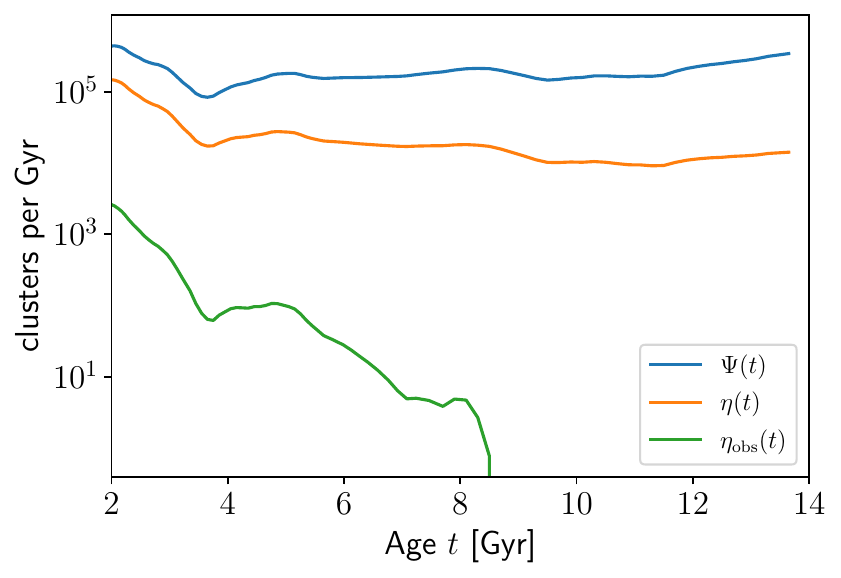}
\caption{CFR and both the complete and observable CAF computed for our model using $M_0=10^5\msun$.}
\label{fig:model}
\end{figure}

Figure~\ref{fig:model} illustrates the different cluster fractions and the model CAF they give rise to.
There, we show the CFR $\Psi (t) = \frac{\Gamma\cdot \sfr}{\langle M \rangle}$, as well as the complete CAF $\eta(t) = f_\mathrm{surv}(t)\Psi (t)$, and the observable CAF $\eta_\mathrm{obs}(t) = f_\mathrm{obs}(t)\Psi (t)$ for SFR and model parameters, as discussed in Sect.~\ref{sec:results}.
The CAF $\eta(t)$ results only from the cluster-forming history and the evolution and dissolution of clusters, while $\eta_\mathrm{obs}(t)$ is also affected by our observational limit, which excludes a part of the cluster population that has not yet dissolved, but is so faint as to not be completely observable.

As the age increases, clusters lose mass, become fainter and dissolve, resulting in a surviving fraction $f_\mathrm{surv}(t)$ that is falling, as seen in the growing gap between $\Psi(t)$ and $\eta(t)$. 
Further, $m_\mathrm{lim}(t)$ rises with age, causing a greater fraction of surviving clusters to fall below the observational limit, as seen in the growing gap between $\eta(t)$ and $\eta_\mathrm{obs}(t)$.

For $t > 8$ Gyr, $m_\mathrm{lim}(t)$ exceeds $10^5\msun$. 
In combination with the MICM and cluster evolution, this results in all surviving clusters older than 9 Gyr falling below the observational limit and not contributing to the observable CAF.

\section{Results and discussion}
\label{sec:results}

\begin{figure}
\center
\includegraphics[scale=0.6]{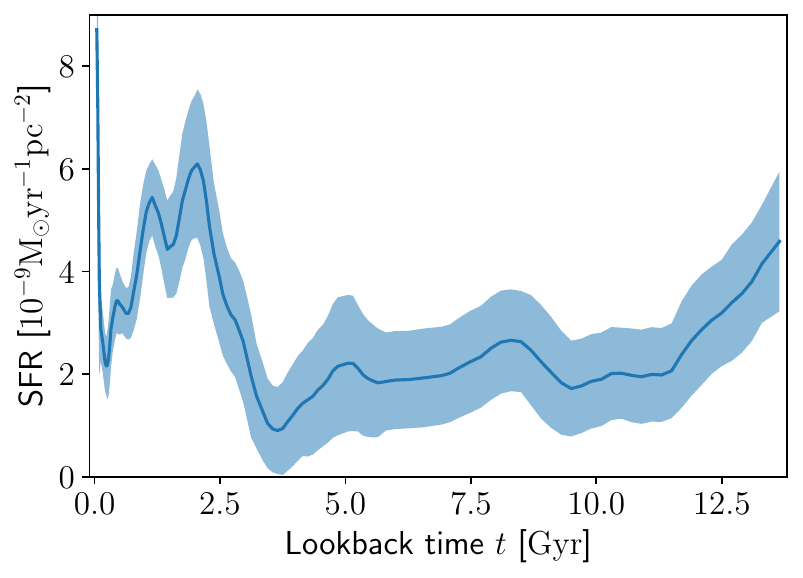}
\caption{LMC SFR including uncertainties derived by \cite{2020A&A...639L...3R}.}
\label{fig:sfr}
\end{figure}

To derive the CAF for our model, we used the SFR constructed by \cite{2020A&A...639L...3R},
which uses photometric data from the Survey of the MAgellanic Stellar History (SMASH) and synthetic colour-magnitude diagram fitting. 
This approach is non-parametric and based on field star data, using models of stellar evolution as the only theoretical input. 
Therefore, we considered their result a reliable estimate of the LMC star-forming history, independent of observations of the LMC cluster system.

Their global LMC SFR with uncertainty band is shown in Fig.~\ref{fig:sfr}.
We note that the SFR shows a burst of star formation at around 2 Gyr, 
as well as comparatively low SFR values between 4 and 11 Gyr. 
Beyond 12 Gyr, the SFR rises again, becoming comparable to the SFR in the recent 1 to 2 Gyr.
In particular, 
\citet{2022MNRAS.513L..40M} link the enhanced LMC SFR in the last 3.5 Gyr to tidal interactions between the LMC and the SMC.

\subsection{MICM relation}
\label{sec:m0_fixed}

As our model is expected to allow for clusters older than 11 Gyr to be observed, we needed to set the MICM relation 
(\ref{eq:max_mass}) appropriately, parametrised by $M_0$ for $\sfr_0 = 1 \msun \mathrm{pc^{-2}Gyr^{-1}}$ and power-law index $\beta$.
Per our review of \cite{2008MNRAS.390..759B}, we set $\beta = 1$ for our model.
According to our model of cluster mass loss, for a cluster to remain observable at ages greater than 11 Gyr, it must have initial mass $M > 4\cdot 10^5\msun$.
Since the SFR generally ranges between 1 and 8 $\msun \mathrm{pc^{-2}Gyr^{-1}}$ for ages $>1$ Gyr, we considered $M_0$ in the range of $5\cdot 10^4\msun$ to $5\cdot 10^5\msun$.

Since the observed masses of ancient clusters reach up to $7.6\cdot 10^5\msun$ and the fact that these clusters will have lost at least half their initial mass just to stellar evolution, the MICM for these clusters must have been at least $1.5\cdot 10^6\msun$, which would require $M_0 \ga 2\cdot 10^5\msun$.
From our discussion below, we find that this constraint indeed coincides with the minimum value of $M_0$ for which an observable population of ancient clusters exists at all.

\begin{figure}
\center
\includegraphics[scale=0.6]{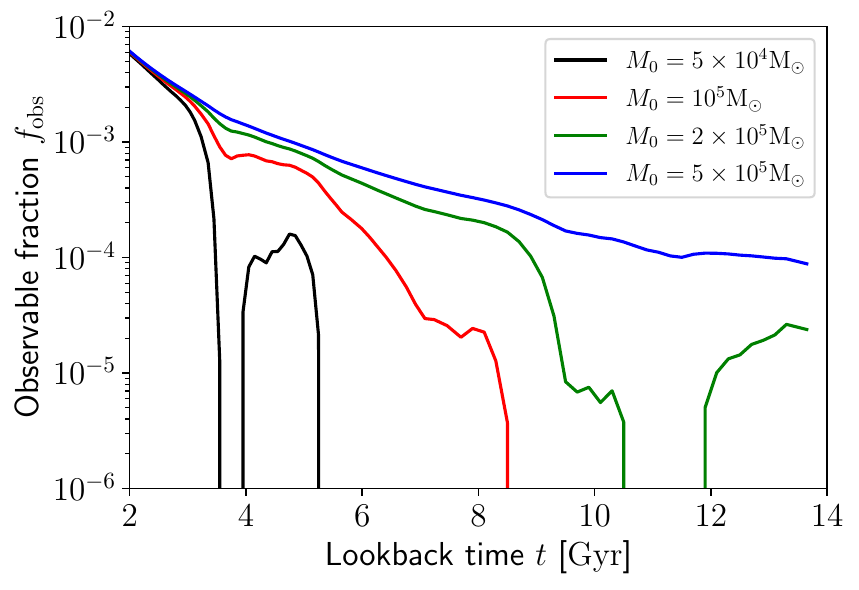}
\caption{Observable cluster fraction in our model as a function of lookback time for $M_\mathrm{max}(t)$ parametrised by $\beta=1$ and $M_0$ in the range $5\cdot 10^4\msun$ to $5\cdot 10^5\msun$.}
\label{fig:fsurv}
\end{figure}

In Fig.~\ref{fig:fsurv}, $f_\mathrm{obs}(t)$ is shown for different values of $M_0$ and age from 2 to 13.8 Gyr.
For some ages, depending on $M_0$, 
$M_\mathrm{max}(t)$ can drop below $M_\mathrm{lim}(R,t)$ for some or all values of $R$, resulting in a sharp decrease of the observable fraction as surviving clusters fall below the mass cut-off. 
In this regime, $f_\mathrm{obs}(t)$ is highly sensitive to the MICM.
Clusters at small orbital radii experience stronger tidal forces and faster dissolution than clusters at larger $R$, so that these clusters fall below $M_\mathrm{lim}(R,t)$ first.
A change in  $M_\mathrm{max}(t)$ causes a change in $f_\mathrm{obs}(t;R)$ for radii where observable clusters remain, and a change in the range of radii, where $f_\mathrm{obs}(t;R) > 0$.
As a result, the smaller $f_\mathrm{obs}(t)$ becomes, the more sensitive it is to $M_\mathrm{max}(t)$ and, thus, to differences in $M_0$.

If the observable fraction is reduced to 0 in this way, a gap in the range of observed cluster ages is the result.
This occurs at lower ages for lower values of $M_0$; for instance, between 10 and 12 Gyr for $M_0 = 2\cdot 10^5\msun$, and above 5--6 Gyr for $M_0 = 5\cdot 10^4\msun$.
For $M_0 = 5\cdot 10^5\msun$, $M_\mathrm{max}(t) > M_\mathrm{lim}(R,t)$ for all ages and radii, so the CAF does not exhibit any gaps.
We note that at earlier ages, clusters have lost less mass and $m_\mathrm{lim}(t)$ is lower. Therefore, $M_\mathrm{lim}(R,t)$ is also lower and the {observable} fraction less sensitive to the precise value of $M_\mathrm{max}(t)$.

\begin{figure}
\center
\includegraphics[scale=0.6]{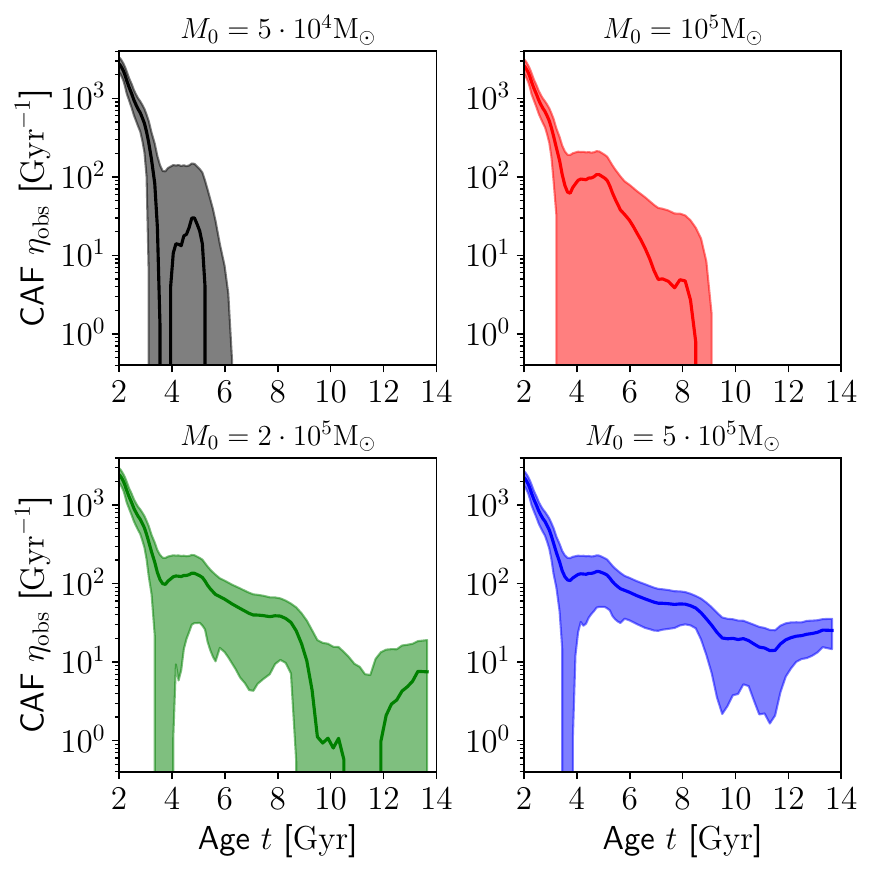}
\caption{Observable CAF $\eta_\mathrm{obs}$ in our model as a function of cluster age for $M_\mathrm{max}(t)$ parametrised by $\beta=1$ and $M_0$ in the range $5\cdot 10^4\msun$ to $5\cdot 10^5\msun$. Shaded areas represent uncertainty bands based on SFR uncertainties.}
\label{fig:eta}
\end{figure}

In Fig.~\ref{fig:eta},
the corresponding resulting CAFs are shown. 
We included uncertainty bands based on the SFR uncertainties of \cite{2020A&A...639L...3R}.
The CAF for $M_0 = 2\cdot 10^5 \msun$ (bottom-left panel)
exhibits an age gap between about 9 and 12 Gyr, which arises from a combination of two effects in this case:
first from a low CFR (fewer clusters formed) and second from the low MICM during periods of low SFR. 
The low MICM results in a greater fraction of surviving clusters falling below the observation threshold.
For $M_0 = 5\cdot 10^4 \msun$ and $M_0=10^5\msun$ (top left and right panels, respectively), 
the CAF does not contain any clusters with age $\ga11$~Gyr and only features an age cut-off around 5--6 and 8--9 Gyr, respectively.
For $M_0 = 5\cdot 10^5\msun$ (bottom right panel), no age gap or age cut-off appears in the CAF.

We also see a sharp decrease of the CAF at ages $\la 4$~Gyr for all values of $M_0$, coinciding with the start of the age gap.
This results from both the SFR and the observable fraction decreasing rapidly for ages 2 to 4 Gyr.

By integrating the model CAF, we derived expected cluster counts of both age gap (4 to 11 Gyr) and ancient clusters (11 to 13.8 Gyr) for a cluster formation efficiency of $\Gamma =0.05$. 
Likewise, integration of the upper and lower limits of the CAF uncertainty band yielded uncertainties of these cluster counts. 
The model for $M_0 = 2\cdot 10^5\msun$, yields an expected number of $9^{+30}_{-9}$ 
ancient clusters, which is compatible with the observed 15 clusters for a wide range of values of $\Gamma$.
However, the model also predicts $340^{+290}_{-280}$ 
age gap clusters above the observational mass limit.
Of the four observed age gap clusters, none lie above their completeness limit.
This model thus cannot explain the observed paucity of clusters in the range of 4 to 11 Gyr.

Similarly, the model for $M_0 = 5\cdot 10^5\msun$ predicts $58^{+31}_{-32}$ 
ancient clusters and $446^{+260}_{-280}$
age gap clusters, thus being compatible with the observed ancient cluster population for lower $\Gamma \approx 0.01$--0.02, but not with the observed age gap.
The corresponding MICM in these two models are also sufficient to reproduce the most massive of the observed ancient clusters.
For both $M_0 = 5\cdot 10^4\msun$ and $M_0 = 10^5\msun$, the model predicts no ancient clusters.
However, the number of expected clusters between ages 4 and 11 Gyr at $23_{-23}^{+174}$ and $175_{-175}^{+277} $
respectively appears compatible with the observations.

\subsection{Varying cluster-forming conditions}
\label{sec:m0_var}

Our results indicate that no single one of our models can reproduce both the observed paucity of clusters in the age gap and the observed population of higher age clusters.
However, the models with lower MICM relation with $M_0 \la 10^5 \msun$ reproduce the paucity of age gap clusters and the models with higher MICM relation with  $M_0 \ga 2\cdot 10^5 \msun$ reproduce the ancient cluster population.

\begin{figure}
\center
\includegraphics[scale=0.6]{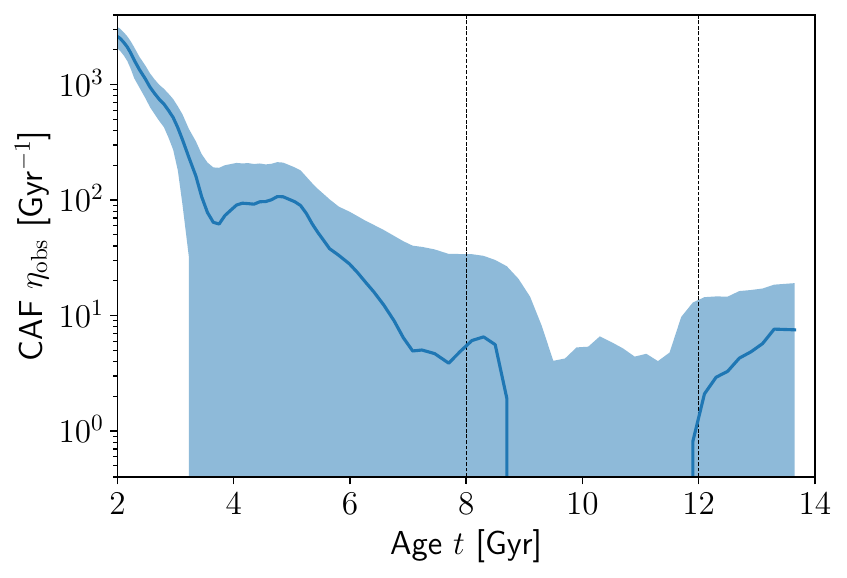}
\caption{CAF $\eta_\mathrm{obs}$ in our model as a function of cluster age for $M_0$ changing linearly from $10^5\msun$ to $2\cdot 10^5\msun$ between ages 8 and 12 Gyr (dotted lines).}
\label{fig:eta_mix}
\end{figure}

If we assume a switch from $M_0 = 2\cdot 10^5 \msun$ to $M_0 = 10^5 \msun$ during lookback times of around 8 to 12~Gyr (e.g.~due to a change to less concentrated cluster formation at lower local SFR intensities), the resulting model will feature both an observed age gap and an ancient cluster population.
In Fig.~\ref{fig:eta_mix},
we show the CAF for a model where $M_0 = 10^5 \msun$ at ages less than 8 Gyr, $M_0 = 2\cdot 10^5 \msun$ at ages above 12 Gyr, and with $M_0$ changing linearly with age for ages 8 to 12 Gyr.
As in the case of the CAF for fixed $M_0 = 2\cdot 10^5 \msun$, it exhibits both an ancient cluster population and an age gap; however, the age gap is wider and the CAF is significantly lower between ages 5 and 10 Gyr.

The model predicts $177^{+294}_{-177}$ clusters between ages 4 and 11 Gyr and $9_{-9}^{+28}$ clusters older than 11 Gyr.
This is compatible with both the observed number of ancient globular clusters and the cluster age gap, showing that such a model of moderately changing $M_0$ is indeed consistent with observations within the SFR uncertainties.

\begin{figure}
\center
\includegraphics[scale=0.6]{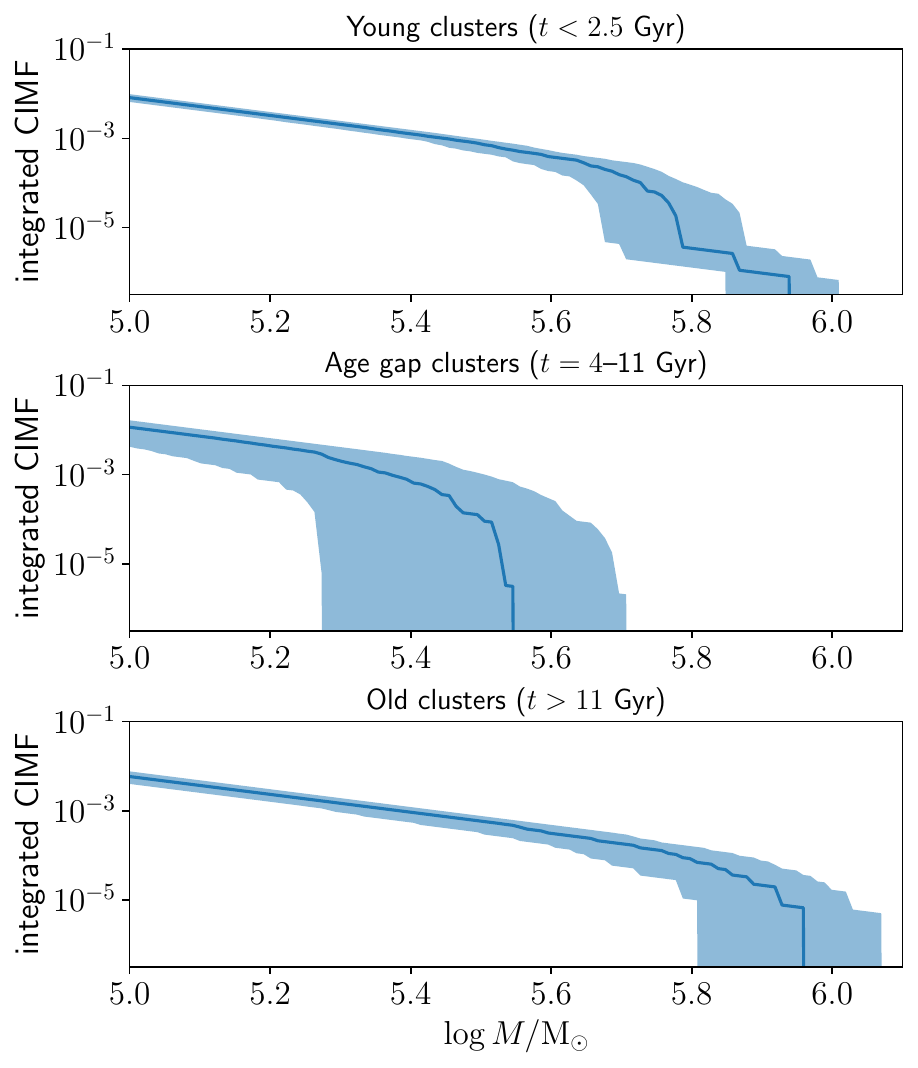}
\caption{Time-integrated CIMFs for recently formed clusters, age gap clusters and old clusters in our model for linearly changing $M_0$.
For $M < 10^5\msun$, the CIMFs follow a power law with $\alpha = 2$.}
\label{fig:cimf}
\end{figure}

In Fig.~\ref{fig:cimf}, we show the integrated CIMF for young clusters formed more recent than 2.5 Gyr ago, for age gap clusters with ages 4--11 Gyr and for old clusters with ages above 11 Gyr.
For an age interval of $[t_1,t_2]$, the time-integrated CIMF $f_{[t_1,t_2]}(M)$ is given by
\begin{equation}
f_{[t_1,t_2]}(M) = \int\limits_{t_1}^{t_2} \Psi(t) f_t(M) \mathrm{d}t,
\end{equation}
where $f_t(M)$ is the CIMF at age $t$, as determined in our model by the SFR and MICM relation.
We note that for $M < 10^5\msun$, all three CIMFs follow the power law $M^{-2}$ and we show only the high-mass end.

Age gap clusters are modelled to have lower MICM than either old clusters or more recently formed clusters.
The changing MICM relation allows for a MICM value in excess of $10^6\msun$ for ancient clusters, despite the SFR not exceeding that of more recent times.
In this sense, the model may be compatible with the observed masses of ancient clusters observed, as discussed also in Sect.~\ref{sec:m0_fixed}. 
The intense present-day burst of LMC star formation is reflected in the low-weight tail for $\log M/\msun > 5.8$ in the integrated CIMF.
This high-mass end of the CIMF for young clusters is not well populated as the present period of high-intensity star formation has only lasted some 100--200 Myr. The model predicts that if star formation in the LMC continues at the current rates, massive clusters possibly up to $10^6\msun$ may form eventually.

\subsection{Limitations}

In our model,
we considered only the global SFR and did not consider spatial variation of the SFR.
However, as we expect the formation of massive clusters to occur in regions of high local SFR
and since it is not clear whether the occurrence of regions with high local SFR can be traced reliably using the global SFR, 
this may be a major weakness of our model that is not easily remedied.
While \cite{2020A&A...639L...3R}
considered the SFR in a spatially resolved way, they used present-day star positions. 
For old field star populations, stars that initially formed in proximity can have significantly migrated away from one another, such that this approach seems unlikely to be able to recover regions of intense star formation in the young LMC, such as the 30 Doradus Nebula in the present-day LMC.

If spatial variations of the SFR were to be taken into account, the global CIMF would then be obtained as a weighted integral of the local CIMF with local MICM depending on the local SFR.
Correspondingly, the upper mass limit of the global CIMF would not be a sharp cut-off, but mirror the distribution of local star-forming intensities, and a time-dependent global MICM relation would arise naturally.
Such a model would utilise the local SFR and tidal conditions to derive local contributions to the CAF, which would then be integrated to obtain the global CAF.
Either way, the SFR acts only as an imperfect indicator of the detailed physical conditions under which clusters formed, whether it be coarse-grained for the global SFR or more fine-grained for a locally resolved SFR.

Another aspect that we did not include was mass loss due to GMC encounters.
As GMC encounters affect low-mass clusters more strongly and high-mass clusters less so \citep[see e.g.~Sect.~4.3.1 of][]{Krumholz2019}, we expect that inclusion of this effect would further reduce the number of age gap cluster relative to the number of ancient clusters.

Finally, while we present a model that reproduces the observed age gap by means of a time-dependent MICM relation in Sect.~\ref{sec:m0_var}, the time dependence we used was chosen based on our previous analysis of fixed MICM to demonstrate the minimum change in cluster formation conditions needed to reproduce the age gap. 
In fact, the high masses of ancient globular clusters point towards a greater change in cluster-forming conditions being preferable. 
Therefore, the model does not represent any kind of best fit estimation, which would require either use of better tracers of cluster formation (e.g. the spatially resolved SFR) and more extensive modelling or further simulations of the LMC evolution, which would come with their own assumptions and limitations.

\section{Conclusion}
\label{sec:conclusion}

It appears plausible that the conditions of star and cluster formation in the LMC have changed over time.
Possibly, cluster formation in the early LMC occurred under higher gas densities and a tighter spatial distribution, as suggested by the high masses of ancient LMC clusters.
Our analysis considering the MICM as a function of global SFR shows that a moderate difference by a factor of 2 to 5 in the MICM-SFR relation between clusters formed between 4 to 10 Gyr ago and clusters formed more than 10 Gyr ago is sufficient to reproduce the observed counts of age gap and ancient clusters.

Based on this premise, our model resolves the apparent discrepancy between observed CAF and SFR at ages above 4 Gyr, thereby providing an explanation for the LMC cluster age gap.
The age gap range from 4 to 11 Gyr coincides with a period of comparatively low SFR.
Consequently, fewer clusters were formed during this time, with initial masses limited by approximately 2$\cdot 10^5\msun$ to 5$\cdot 10^5\msun$. 

Due to cluster mass loss and stellar evolution, the majority of
these age gap clusters have dissolved and the remainder have
become so faint 
that they fall below present observational limits. 
The four age gap clusters that have been observed are all of low mass, with the most massive reaching $5\cdot 10^3\msun$.
It is the prediction of our model that there is no undetected population of massive age-gap clusters and that future detections of age-gap clusters will capture similarly low-mass clusters.
An enhanced disruption of age gap clusters (e.g. due to tidal interactions of the LMC) is not needed in our model.

The SFR for ages $>12$ Gyr exceeds the SFR during the age gap range. 
In our model, this period of higher SFR corresponds to a greater number of clusters as well as more massive clusters being formed.
These massive clusters with initial masses on the order of $10^6\msun$ now comprise the observed population of massive ancient clusters with ages exceeding 11 Gyr.

Notably, our results suggest that a galaxy's star-forming history can be used to estimate its cluster-forming history, with the conditions of cluster formation (at least at the high-mass end of the CIMF) being modelled by a function of the SFR that changes very slowly over time.
This reinforces the picture of globular cluster formation being linked to a sufficiently high star-forming density, which may arise due to any confluence of factors, rather to any specific formation mechanism,  

In summary, the existence of old globular clusters in a galaxy such as the LMC is a consequence of the young galaxy's high gas content enabling intense star formation and the birth of massive star clusters early on in its existence. As the initial burst of star formation subsides and the galaxy enters a period of lower SFR at lower densities, only less massive clusters are formed, most of which have not survived to the present day. 
The surviving clusters of the low SFR period have low remaining masses and are faint, making them difficult to observe.
Younger clusters retain more mass as well as a population of higher mass main sequence stars. They can be numerous and luminous enough to be more easily detectable, even if no further increase of the SFR occurs.
Therefore, the observation of a young and an old population of clusters separated by an age gap would not be uncommon or surprising. Rather, observing this phenomenon would be the result of an observational sensitivity that has the capacity to detect both young, bright clusters and old globular clusters, without being sufficient to detect less massive, evolved clusters at all intermediate ages.

\begin{acknowledgements}
We thank the referee for the thorough reading of the manuscript and helpful suggestions to improve it.

We thank T.~Ruiz-Lara for providing us his LMC SFR data sets.

Data for reproducing the figures and analysis in this work are available upon request to the first author.

Analysis was done in Python using the \verb+numpy+\footnote{\url{https://numpy.org}} library, and  
the figures were produced using the \verb+matplotlib+\footnote{\url{https://matplotlib.org}} library.

\end{acknowledgements}

\bibliographystyle{aa}
\bibliography{paper}

\begin{thebibliography}{30}
\expandafter\ifx\csname natexlab\endcsname\relax\def\natexlab#1{#1}\fi

\bibitem[{{Bastian}(2008)}]{2008MNRAS.390..759B}
{Bastian}, N. 2008, \mnras, 390, 759

\bibitem[{{Baumgardt} {et~al.}(2013){Baumgardt}, {Parmentier}, {Anders}, \&
  {Grebel}}]{2013MNRAS.430..676B}
{Baumgardt}, H., {Parmentier}, G., {Anders}, P., \& {Grebel}, E.~K. 2013,
  \mnras, 430, 676

\bibitem[{{Bekki} {et~al.}(2004){Bekki}, {Couch}, {Beasley}, {Forbes}, {Chiba},
  \& {Da Costa}}]{bekkietal2004}
{Bekki}, K., {Couch}, W.~J., {Beasley}, M.~A., {et~al.} 2004, \apjl, 610, L93

\bibitem[{{Berek} {et~al.}(2023){Berek}, {Reina-Campos}, {Eadie}, \&
  {Sills}}]{bereketal2023}
{Berek}, S.~C., {Reina-Campos}, M., {Eadie}, G., \& {Sills}, A. 2023, \mnras,
  525, 1902

\bibitem[{{Da Costa}(1991)}]{dc1991}
{Da Costa}, G.~S. 1991, in IAU Symposium, Vol. 148, The Magellanic Clouds, ed.
  R.~{Haynes} \& D.~{Milne}, 183

\bibitem[{{Ferreira} {et~al.}(2025){Ferreira}, {Dias}, {Santos}, {Maia},
  {Bica}, {Kerber}, {Armond}, {Quint}, {Oliveira}, {Souza},
  {Fern{\'a}ndez-Trincado}, \& {Katime Santrich}}]{ferreiraetal2025}
{Ferreira}, B. P.~L., {Dias}, B., {Santos}, J. F.~C., {et~al.} 2025, \aap, 695,
  L9

\bibitem[{{Gatto} {et~al.}(2020){Gatto}, {Ripepi}, {Bellazzini}, {Cignoni},
  {Cioni}, {Dall'Ora}, {Longo}, {Marconi}, {Schipani}, \&
  {Tosi}}]{gattoetal2020}
{Gatto}, M., {Ripepi}, V., {Bellazzini}, M., {et~al.} 2020, \mnras, 499, 4114

\bibitem[{{Gatto} {et~al.}(2024){Gatto}, {Ripepi}, {Bellazzini}, {Tosi},
  {Cignoni}, {Tortora}, {Marconi}, {Dall'Ora}, {Cioni}, {Musella}, {Schipani},
  \& {Spavone}}]{gattoetal2024}
{Gatto}, M., {Ripepi}, V., {Bellazzini}, M., {et~al.} 2024, \aap, 690, A164

\bibitem[{{Gatto} {et~al.}(2022){Gatto}, {Ripepi}, {Bellazzini}, {Tosi},
  {Tortora}, {Cignoni}, {Dall'Ora}, {Cioni}, {Cusano}, {Longo}, {Marconi},
  {Musella}, {Schipani}, \& {Spavone}}]{gattoetal2022}
{Gatto}, M., {Ripepi}, V., {Bellazzini}, M., {et~al.} 2022, \aap, 664, L12

\bibitem[{{Geisler} {et~al.}(1997){Geisler}, {Bica}, {Dottori}, {Claria},
  {Piatti}, \& {Santos}}]{getal97}
{Geisler}, D., {Bica}, E., {Dottori}, H., {et~al.} 1997, \aj, 114, 1920

\bibitem[{{Harris} \& {Zaritsky}(2009)}]{2009AJ....138.1243H}
{Harris}, J. \& {Zaritsky}, D. 2009, \aj, 138, 1243

\bibitem[{{Kruijssen} \& {Cooper}(2012)}]{2012MNRAS.420..340K}
{Kruijssen}, J.~M.~D. \& {Cooper}, A.~P. 2012, \mnras, 420, 340

\bibitem[{{Krumholz} {et~al.}(2019){Krumholz}, {McKee}, \&
  {Bland-Hawthorn}}]{Krumholz2019}
{Krumholz}, M.~R., {McKee}, C.~F., \& {Bland-Hawthorn}, J. 2019, \araa, 57, 227

\bibitem[{{Lamers} {et~al.}(2010){Lamers}, {Baumgardt}, \&
  {Gieles}}]{Lamers2010}
{Lamers}, H. J.~G.~L.~M., {Baumgardt}, H., \& {Gieles}, M. 2010, \mnras, 409,
  305

\bibitem[{{Li} {et~al.}(2018){Li}, {Yanny}, \& {Wu}}]{lietal2018}
{Li}, G.-W., {Yanny}, B., \& {Wu}, Y. 2018, \apj, 869, 122

\bibitem[{{Maia} {et~al.}(2014){Maia}, {Piatti}, \& {Santos}}]{metal14}
{Maia}, F.~F.~S., {Piatti}, A.~E., \& {Santos}, J.~F.~C. 2014, \mnras, 437,
  2005

\bibitem[{{Massana} {et~al.}(2022){Massana}, {Ruiz-Lara}, {No{\"e}l},
  {Gallart}, {Nidever}, {Choi}, {Sakowska}, {Besla}, {Olsen}, {Monelli},
  {Dorta}, {Stringfellow}, {Cassisi}, {Bernard}, {Zaritsky}, {Cioni},
  {Monachesi}, {van der Marel}, {de Boer}, \& {Walker}}]{2022MNRAS.513L..40M}
{Massana}, P., {Ruiz-Lara}, T., {No{\"e}l}, N.~E.~D., {et~al.} 2022, \mnras,
  513, L40

\bibitem[{{Mateo} {et~al.}(1986){Mateo}, {Hodge}, \&
  {Schommer}}]{mateoetal1986}
{Mateo}, M., {Hodge}, P., \& {Schommer}, R.~A. 1986, \apj, 311, 113

\bibitem[{{Pagel} \& {Tautvaisiene}(1998)}]{pt1998}
{Pagel}, B.~E.~J. \& {Tautvaisiene}, G. 1998, \mnras, 299, 535

\bibitem[{{Piatti}(2014)}]{2014MNRAS.437.1646P}
{Piatti}, A.~E. 2014, \mnras, 437, 1646

\bibitem[{{Piatti}(2021)}]{piatti2021d}
{Piatti}, A.~E. 2021, \aj, 161, 199

\bibitem[{{Piatti}(2022)}]{piatti2022c}
{Piatti}, A.~E. 2022, \mnras, 511, L72

\bibitem[{{Piatti}(2025{\natexlab{a}})}]{piatti2025c}
{Piatti}, A.~E. 2025{\natexlab{a}}, \mnras, 541, 3763

\bibitem[{{Piatti}(2025{\natexlab{b}})}]{2025MNRAS.537.1586P}
{Piatti}, A.~E. 2025{\natexlab{b}}, \mnras, 537, 1586

\bibitem[{{Piatti} {et~al.}(2019){Piatti}, {Alfaro}, \&
  {Cantat-Gaudin}}]{piattietal2019}
{Piatti}, A.~E., {Alfaro}, E.~J., \& {Cantat-Gaudin}, T. 2019, \mnras, 484, L19

\bibitem[{{Piatti} {et~al.}(2025){Piatti}, {Illesca}, {Chiarpotti}, \&
  {Butr{\'o}n}}]{piattietal2025}
{Piatti}, A.~E., {Illesca}, D.~M.~F., {Chiarpotti}, M., \& {Butr{\'o}n}, R.
  2025, \aap, 702, A108

\bibitem[{{Piatti} \& {Mackey}(2018)}]{pm2018}
{Piatti}, A.~E. \& {Mackey}, A.~D. 2018, \mnras, 478, 2164

\bibitem[{{Pieres} {et~al.}(2016){Pieres}, {Santiago}, {Balbinot}, {Luque},
  {Queiroz}, {da Costa}, {Maia}, {Drlica-Wagner}, {Roodman}, {Abbott}, {Allam},
  {Benoit-L{\'e}vy}, {Bertin}, {Brooks}, {Buckley-Geer}, {Burke}, {Carnero
  Rosell}, {Carrasco Kind}, {Carretero}, {Cunha}, {Desai}, {Diehl}, {Eifler},
  {Finley}, {Flaugher}, {Fosalba}, {Frieman}, {Gerdes}, {Gruen}, {Gruendl},
  {Gutierrez}, {Honscheid}, {James}, {Kuehn}, {Kuropatkin}, {Lahav}, {Li},
  {Marshall}, {Martini}, {Miller}, {Miquel}, {Nichol}, {Nord}, {Ogando},
  {Plazas}, {Romer}, {Sanchez}, {Scarpine}, {Schubnell}, {Sevilla-Noarbe},
  {Smith}, {Soares-Santos}, {Sobreira}, {Suchyta}, {Swanson}, {Tarle},
  {Thaler}, {Thomas}, {Tucker}, \& {Walker}}]{pieresetal2016}
{Pieres}, A., {Santiago}, B., {Balbinot}, E., {et~al.} 2016, \mnras, 461, 519

\bibitem[{{Ruiz-Lara} {et~al.}(2020){Ruiz-Lara}, {Gallart}, {Monelli},
  {Nidever}, {Dorta}, {Choi}, {Olsen}, {Besla}, {Bernard}, {Cassisi},
  {Massana}, {No{\"e}l}, {P{\'e}rez}, {Rusakov}, {Cioni}, {Majewski}, {van der
  Marel}, {Mart{\'\i}nez-Delgado}, {Monachesi}, {Monteagudo}, {Mu{\~n}oz},
  {Stringfellow}, {Surot}, {Vivas}, {Walker}, \&
  {Zaritsky}}]{2020A&A...639L...3R}
{Ruiz-Lara}, T., {Gallart}, C., {Monelli}, M., {et~al.} 2020, \aap, 639, L3

\bibitem[{{Zhou} {et~al.}(2025){Zhou}, {Kroupa}, \& {Dib}}]{zhouetal2025}
{Zhou}, J.~W., {Kroupa}, P., \& {Dib}, S. 2025, \mnras, 541, 1276

\end{thebibliography}

\end{document}